\documentclass{amsart}
\usepackage[latin1]{inputenc}
\usepackage{amsmath}
\usepackage{amsfonts}
\usepackage{amssymb}
\usepackage{graphicx}
\usepackage{rotating}

\newcommand{\mc}{\mathcal}
\newcommand{\mbb}{\mathbb}
\newcommand{\be}{\begin{equation*}}
\newcommand{\ee}{\end{equation*}}
\newcommand{\bee}{\begin{equation}}
\newcommand{\eee}{\end{equation}}

\newcommand{\ind}{\mathrm{ind}}
\newcommand{\res}{\mathrm{res}}

\newtheorem{Theorem}{Theorem}[section]
\newtheorem{Lemma}[Theorem]{Lemma}
\newtheorem{Proposition}[Theorem]{Proposition}

\newenvironment{Proof}[1][Proof.]{\begin{trivlist}
\item[\hskip \labelsep {\bfseries #1}]}{\end{trivlist}}

\newenvironment{Remark}[1][Remark.]{\begin{trivlist}
\item[\hskip \labelsep {\bfseries #1}]}{\end{trivlist}}

\begin{document}

\title[Discrete analogues of Dirac's magnetic monopole]{Discrete analogues of Dirac's magnetic monopole and binary polyhedral groups}

\author{G.M. Kemp}
\address{Department of
Mathematical Sciences,
          Loughborough University,
Loughborough,
          Leicestershire, LE11 3TU, UK}
       \email{G.Kemp@lboro.ac.uk}

\author{A.P. Veselov}
\address{Department of
Mathematical Sciences,
          Loughborough University,
Loughborough,
          Leicestershire, LE11 3TU, UK
          and
Moscow State University, Moscow 119899, Russia}

\email{A.P.Veselov@lboro.ac.uk}

\maketitle

{\begin{center}\normalsize\it
     Dedicated to S. P. Novikov
     on his 75-th birthday
   \end{center}}
   
   \bigskip

{\small  {\bf Abstract.} We introduce some discrete analogues of the Dirac magnetic monopole on a unit sphere $S^2$ and explain how to compute the corresponding spectrum using the representation theory of finite groups. The main examples are certain magnetic Laplacians on the regular polyhedral graphs, coming from induced representations of the binary polyhedral groups.}

\section{Introduction}

In 1931 Dirac introduced his famous magnetic monopole in \cite{Dirac} and showed that the corresponding magnetic charge $q$ should be quantized: $q \in \mathbb Z.$  Its version on a unit sphere was analysed in 1976 by
Wu and Yang \cite{WY}, who showed that the wavefunctions are the sections of a complex line bundle $L$ over $S^2$ and found the spectrum explicitly.
Since the magnetic charge is the first Chern class $c_1(L)=q$ of the bundle $L$, which must be an integer, this gives also a geometric interpretation of Dirac's quantization condition.  

In \cite{KV} we proposed a simple derivation of the spectrum of the Dirac magnetic monopole on a unit sphere $S^2$ based on geometric quantization and the Frobenius reciprocity formula. The initial point for us was the calculation by S.P. Novikov and Schmelzer \cite{NS} of the canonical symplectic structure on the coadjoint orbits of the Euclidean group $E(3)$ of motions  of $\mathbb{E}^3,$ which showed the relation with the classical Dirac monopole. 
We showed that Novikov--Schmelzer variables have a natural quantum version as covariant derivatives acting on the space of sections $\Gamma(L)$ of the corresponding line bundle $L.$ 

Algebraically the space $\Gamma(L)$ can be viewed as the representation of $SU(2)$ induced from the representation of $U(1) \subset SU(2)$ given by $z \rightarrow z^{q}, z \in U(1)$ (see \cite{Bott}). This space can be decomposed as an $SU(2)$-module using  the classical Frobenius reciprocity formula \cite{FH}. Since the quantum Hamiltonian is essentially the quadratic Casimir operator this leads to the Wu-Yang formula for the Dirac monopole spectrum (see \cite{KV} and the next section).

In this paper we propose a class of discrete versions of Dirac magnetic monopole on a sphere 
by taking this algebraic point of view and extending it to the discrete groups. The main idea is the following.

First we replace the unit sphere $S^2=SU(2)/U(1)$ by the quotient space $X=G/H$, 
where $G$ is a finite group and $H$ is a subgroup. The main question is what  
should play the role of the metric and Hamiltonian. We propose to choose a suitable element $C$ 
from the centre of the group ring $\mathbb Z[G]$ of group $G,$ which is the sum of elements from a conjugacy class of $G.$ 
For such an element we can construct a graph 
$\Gamma=\Gamma(G,H,C)$ with the  set of vertices identified with $X.$
The main conditions on $G,H,C$ are that the graph $\Gamma$ is planar and the action of $G$ 
on the edges of $\Gamma$ is transitive.

The magnetic field corresponds to the choice of a character $\chi:H\rightarrow U(1)$ 
and the Hilbert space $V$ of the system is the induced representation $V=\mathrm{ind}_H^G(\chi).$
The action of $C$ on $V$ determines in a natural way certain magnetic Laplacian \cite{LL}
on the graph $\Gamma,$ which is considered as the corresponding discrete Dirac magnetic monopole.
Their spectrum can be computed using the Frobenius reciprocity formula similarly to the usual case.

The main examples are given by the binary polyhedral groups $G \subset SU(2)$ with a proper choice of $H$ and $C,$ 
which give magnetic Laplacians on all regular polyhedral graphs of platonic solids with the exception of dodecahedron.

We should mention that there exists a different approach to the discretisation based on triangulation of manifolds, 
which was initiated by S.P. Novikov \cite{N, N2, N3}. Although it is based on very different ideas 
this was a part of the motivation for our work.

\section{Geometric quantisation of Dirac magnetic monopole}

We start with classical Dirac magnetic monopole and its geometric quantization, mainly following our paper \cite{KV}. 

Let $e(3)$ be the Lie algebra  of the Euclidean group $E(3)$ of motions of $\mathbb{E}^3.$ It has the basis $l_i, p_i, \, i=1,2,3$, where $p_i$ and $l_i$ are generators of translations and rotations respectively. 

The dual space $e(3)^*$ has the canonical Lie--Poisson structure
\begin{equation*}
\left\{l_i, l_j\right\} = \epsilon_{ijk} l_k, \ \  \left\{l_i, p_j\right\} = \epsilon_{ijk} p_k, \ \  \left\{p_i, p_j \right\}=0.
\end{equation*}
Its symplectic leaves are the coadjoint orbits of $E(3)$ determined by
\begin{equation*}
(p,p)=R^2, \qquad (l,p)=\alpha R.
\end{equation*} 
Novikov and Schmelzer \cite{NS} introduced the variables
\begin{equation}
\label{NS}
 \sigma_i=l_i-\frac{\alpha}{R}p_i, \quad i=1,2,3,
 \end{equation}
which allows us to identify the coadjoint orbits with $T^*S^2:$ 
\begin{equation*}
(p,p)=R^2, \qquad (\sigma,p)=0.
\end{equation*}
Assume for convenience from now on that $R=1.$

In the new variables the Poisson brackets are
\begin{equation}
\left\{\sigma_i, \sigma_j\right\} = \epsilon_{ijk}\left(\sigma_k -\alpha p_k \right), \ \  \left\{\sigma_i, p_j\right\} = \epsilon_{ijk} p_k, \ \  \left\{p_i, p_j \right\}=0 \label{eq:PBs}
\end{equation}
and the corresponding canonical symplectic form is
\begin{equation}
\omega=\, \mathrm{d} P \wedge \, \mathrm{d} Q +  \alpha  \, \mathrm{d} S, \label{eq:symp}
\end{equation}
where $\, \mathrm{d} P \wedge \, \mathrm{d} Q$ is the standard symplectic form on $T^* S^2$ and $\, \mathrm{d} S$ is the area form on $S^2$ (see \cite{NS}). 
As it was pointed out in \cite{NS} the second term corresponds to the magnetic field of the (non-quantized) Dirac monopole:
$
\mathcal H= \alpha \, \mathrm{d} S.
$

Consider now a complex line bundle over $S^2$ with a $U(1)$-connection having the curvature form
\begin{equation*}
\mathcal  R= i \mathcal  H = i \alpha \, \mathrm{d} S
\end{equation*} 
motivated by geometric quantization (see e.g. \cite{Hu}).
Since the first Chern class of the bundle must be an integer we have 
$$q=\frac{1}{2\pi i} \int_{S^2} \mathcal R = \frac{1}{2\pi} \int_{S^2} \alpha \mathrm{d} S=2 \alpha \in \mathbf Z,$$
which is precisely Dirac's quantization condition. 

Let $$X_1=x_3 \partial_2 - x_2 \partial_3,\ X_2= x_1 \partial_3 - x_3 \partial_1,\ X_3= x_2 \partial_1 - x_1 \partial_2$$ be the vector fields generating rotations of $S^2$ given by $x_1^2+x_2^2+x_3^2=1$ and $\nabla_{X_j}$ be the corresponding covariant derivatives.
The claim \cite{KV} is that
$$\hat \nabla_j:=i \nabla_{X_j}$$ and the operators $\hat x_j$ of multiplication by  $x_j$ satisfy the commutation relations
$$[\hat \nabla_k, \hat \nabla_l] = i \epsilon_{klm}(\hat \nabla_m - \alpha \hat x_m)$$
and thus can be considered as quantization of Novikov-Schmelzer variables.
The quantum versions of the original variables
\begin{equation}
\hat l_j = \hat \nabla_j + \alpha x_j,
\label{mam}
\end{equation}
satisfy the standard angular momentum relations 
\begin{equation*}
[\hat l_k, \hat l_m] = i\epsilon_{kmn}\hat l_n,
\end{equation*} 
and coincide with Fierz's modification of the angular momentum 
in the presence of the Dirac magnetic monopole \cite{Fierz}. 

The quantum Hamiltonian of the Dirac monopole can be written 
in terms of Novikov-Schmelzer operators as
\begin{equation*}
\hat H=\hat \sigma^2
\end{equation*}  
or, equivalently,  in terms of magnetic angular momentum $\hat l$ as
\begin{equation*}
\hat H=\hat l^2-\alpha^2=\hat l^2-\frac{1}{4} q^2.
\end{equation*}
Since the operator $\hat l^2$ is a Casimir operator for $SU(2)$ it 
 acts on every irreducible representation of $SU(2)$ as a scalar.
 More precisely, if $V_k$ is the irreducible representation with highest weight $k \in \mathbf Z_{\geq 0}$
 then 
\begin{equation}
\hat l^2|_{V_k} = \frac{1}{4}k(k+2)I, \label{eq:Lsquared}
\end{equation} 
see e.g. \cite{FH}. 

Now the key observation is that the Hilbert space 
of the quantum system, which is the space of sections of the line bundle $L_q$ over $S^2=SU(2)/U(1)$ with first Chern class $q$, can be interpreted algebraically as the induced representation 
\begin{equation*}
\Gamma(L_q)=\mathrm{ind}_{U(1)}^{SU(2)}\left(W_q\right),
\end{equation*}
where $W_q$ is representation of $U(1)$ given by 
$e^{i \theta} \mapsto e^{i q \theta}, \,\, q \in \mathbf Z$
(see \cite{Bott, KV}). Thus to compute the spectrum of $\hat H$ one needs only to know the decomposition of $\mathrm{ind}_{U(1)}^{SU(2)}\left(W_q\right)$ into irreducible representations of $SU(2).$ 

This can be done using the classical {\it Frobenius reciprocity formula} 
\begin{equation}
\left\langle V,\mathrm{ind}_{H}^{G}\left(W\right) \right\rangle_{G}=\left\langle W, \mathrm{res}_{G}^{H}(V)\right\rangle_{H}. \label{eq:frobG}
\end{equation}
Here $G$ is a group, $H$ is its subgroup, $V$ and $W$ are the irreducible representations of $G$ and $H$ respectively, $\mathrm{ind}_{H}^{G}(W)$ is the representation of $G$ induced from $W$,
$\mathrm{res}_{G}^{H}(V)$ is the restriction of the representation $V$ to the subgroup $H$ and the brackets denote the multiplicity of the first representation entering into the second one (see \cite{FH}).

In our case this gives (see \cite{KV})
\begin{equation}
\Gamma(L_q)=\mathrm{ind}_{U(1)}^{SU(2)}\left(W_q\right) =\bigoplus_{l\in\mathbb{Z}_{\geq0}}{V_{2l+|q|}}. \label{eq:decomp}
\end{equation}
The space $V_{2l+|q|}$ has dimension $2l + |q| +1$, and for $\psi \in V_{2l+|q|}$, the operator $\hat H$ acts as
$$
\hat H\psi=(\hat l^2-\frac{1}{4} q^2)\psi=\left[\frac{1}{4}\left(2l+|q|\right) \left(2l+|q|+2\right) - \frac{1}{4} q^2 \right]\psi=  \left[l(l+1)+|q|\left(l+\frac{1}{2}\right) \right]\psi
$$
Thus the spectrum of the Dirac magnetic monopole with magnetic charge $q$ on the unit sphere $S^2$  is given by
\begin{equation}
\lambda = \left[l(l+1)+|q|\left(l+\frac{1}{2}\right)\right], l=0,1,2,\dots \ \text{with degeneracy } \  2l+|q|+1 \label{eq:spec}
\end{equation}
in agreement with Wu and Yang.  

\section{Magnetic fields on graphs}

In looking for a discrete analogue of the Dirac magnetic monopole, it is natural to look at the theory of magnetic fields on graphs.  We present a brief review of the basics of this theory here, referring to \cite{LL} and \cite{CdV} for justifications and background.  

Recall that a graph $\Gamma= \Gamma(\mathcal V,\mathcal E)$ is a collection of vertices $\mathcal V$ joined by a set of edges $\mathcal E$.  We denote arbitrary vertices by Roman letters $x, y, \ldots \in \mathcal V$ and an unoriented edge is written as $\{x,y\}$, with its two orientations being $[x,y]$ and $[y,x]$.  

For simplicity, we will restrict to graphs with no loops (self-connections) or multiple edges.  The graphs that we will construct in the next section will all be finite and of a special type, satisfying the following regularity condition.  

A graph $\Lambda$ is said to  be \textit {$d$-regular} if the number of edges joined to each vertex is $d$ and there are no multiple edges or self-connections allowed.  These graphs are special and have the property that locally  each vertex looks the same, and thus can be thought of as an analogue of manifolds.  

The space of functions on a finite graph $\Lambda$ is denoted by
\be
C(\mathcal V) := \left\{f: \mathcal V \rightarrow \mathbb{C} \right\}.
\ee
It is a $|\mathcal V|$-dimensional space, where $|\mathcal V|$ is the number of vertices of $\Gamma$, with a natural basis $e_x, \, x\in \mathcal V$ given by 
$e_x(x)=1$ and $e_x(y)=0$ for all $y \neq x.$ By slightly abusing notation, we will not distinguish the operators acting in $C(\mathcal V)$ and their matrices in the basis $e_x.$

The \textit{adjacency matrix} $A$ of a graph $\Gamma$ records which vertices  are linked by edges: it is the $|\mathcal V| \times |\mathcal V|$ matrix indexed by the vertices of $\Gamma$ with the property that $A_{xy}=1$ if $[x,y] \in \mathcal E$ and $A_{xy} = 0$ otherwise.  
The adjacency matrix acts on functions as
\be
(Af)(x)= \sum_{y \sim x}{f(y)},
\ee
where the sum is over all vertices $y$ that are linked to $x$ by an edge.

The graph $\Gamma$ is undirected if $A=A^t$.  The adjacency matrix can be thought of as defining a discrete metric on $\Gamma$ with the entries of $A^k$ counting the random walks in $\Gamma$ of length $k$.    

The \textit{Laplacian} $\mathcal L$ of $\Gamma$ is defined as the operator acting on functions as
\be
(\mathcal Lf)(x)= \sum_{y \sim x}{\left(f(x)-f(y)\right)},
\ee
where the sum is over all vertices $y$ that are linked to $x$ by an edge. The adjacency matrix $A$ may be obtained from the matrix of the Laplacian $\mathcal L$ by forgetting the diagonal terms and changing signs.

Thus, we have two natural operators $A$ and $\mathcal L$ acting on $C(\mathcal V)$, which in general have two different eigenvalue problems (see \cite{LL} for a discussion).  However, for $d$-regular graphs they are essentially equivalent since 
$\mathcal L = d I - A,$
where $I$ is the $|\mathcal V| \times |\mathcal V|$ identity matrix.  

We now explain how to introduce a magnetic field on a graph, following \cite{CdV,LL}.   

A \textit{magnetic potential} $\mc{A}$ on an undirected graph $\Gamma$ with no multiple edges or loops is given by associating to each edge $[x,y]$ an element $\exp[i \alpha_{xy}] \in U(1)$ such that  $\alpha_{xy} \equiv -\alpha_{yx}\in \mathbb{R} \mod 2 \pi$.  The corresponding \textit{magnetic adjacency matrix} $A=A_\mc{A}$ has the matrix elements $A_{xy}= \exp[i \alpha_{xy}]$ for $[x,y] \in \mathcal E$.  

Given a magnetic potential $\mc{A}$ on $\Gamma$, define  the \textit{magnetic Laplacian} $\mathcal L_{\mc{A}}$ as an operator acting on $C(\mathcal V)$
\bee
(\mathcal L_{\mc{A}}f)(x) = \sum_{y \sim x}{\left[f(x)-\exp[i \alpha_{xy}]f(y)\right]}, \label{eq:maglap}
\eee
with the summation being taken over all vertices $y$ that are joined to $x$ by an edge. 
Both the magnetic Laplacian and the magnetic adjacency matrices are Hermitian since $\exp[i\alpha_{xy}]=\overline{\exp[i\alpha_{yx}]}$.   

A \textit{gauge transformation} $U$ is defined by 
\be
(Uf)(x) = \exp[i \sigma_x] f(x),
\ee
where $\sigma_x \in \mathbb{R}$. It acts on magnetic potentials by $\alpha_{xy} \mapsto \alpha_{xy}+ \sigma_y - \sigma_x$
and on the magnetic Laplacians by 
\be
\mathcal L_{\mc{A}} \mapsto \overline{U}^t \mathcal L_{\mc{A}} U=U^{-1} \mathcal L_{\mc{A}} U,
\ee
where $U_{xy}=\exp[i \sigma_x] \delta_{xy}$. Gauge transformations do not change the spectra of $A_\mc{A}$ and $\mathcal L_{\mc{A}}$. 

Given a magnetic potential $\mc{A}$ and a cycle $\gamma= [x_0,x_1] + [x_1,x_2]+ \ldots +[x_{n-1},x_n]$ (with $x_n=x_0$) on $\Gamma$, one can define the \textit{magnetic flux} $\Phi(\gamma)$ through $\gamma$ by
\be
\Phi(\gamma)= \arg\left( \prod_{i=1}^n \exp[i \alpha_{x_{i-1} x_{i}}]\right)  \equiv \sum_{i=1}^n \alpha_{x_{i-1} x_{i}} \in \mathbb{R} \mod 2 \pi. \label{eq:mflux}
\ee
We give here, without proof, two basic results explaining how the flux through each cycle affects the spectrum.  

\begin{Proposition} \cite{LL}
Let $\mc{A}$ and $\mc{A'}$ be two magnetic potentials on a graph $\Gamma$, such that the corresponding fluxes through each cycle are equal. Then $\mc{A}$ and $\mc{A'}$ are gauge equivalent and consequently the corresponding magnetic adjacency matrices and magnetic Laplacians are isospectral.   
\end{Proposition}

\begin{Proposition} \cite{CdV}
For  a connected graph $\Gamma$, zero is an eigenvalue of a magnetic Laplacian if and only if the flux through each cycle is 0, so the corresponding magnetic potential is gauge equivalent to zero.
\end{Proposition}

\section{Magnetic fields on regular graphs related to finite groups}
In this section we define a class of regular graphs related to finite groups and explain how to define on them an invariant magnetic field.     

Let $G$ be a finite group and define a \textit{Casimir element} of $G$ to be any element belonging to the centre of the group ring $\mathbb{Z}[G]$.  
The terminology is explained by the analogy with Lie group theory, where the Casimir elements belong to the centre of the corresponding universal enveloping algebra.  In particular, for a semisimple Lie group $G$ a Casimir element of the second order defines both metric and Laplace-Beltrami operator on the homogeneous spaces of $G$.  We are going to describe now some special Casimir elements for a finite group, which can be used in a similar way to define certain graphs and Laplacians on it.

It is easy to see that given any conjugacy class $[c]$ of $G$, we can define the Casimir element $C$ corresponding to $[c]$, by taking the formal sum of  each element in $[c]$
\bee
C := \sum_{c \in [c]}{c} \label{eq:cassum}. 
\eee
These elements generate the centre of the group ring $\mathbb{Z}[G]$.

Since a Casimir element $C$ commutes with all elements of $G$, by Schur's lemma it acts on each irreducible representation $V$ of $G$ by  multiplication by a scalar $\lambda_V \in \mathbb{C}$.  If $C$ is of the form \eqref{eq:cassum}, then $\lambda_V$ is given by 
\bee
\lambda_V = \frac{\chi_V(c) \cdot |[c]|}{\dim V}, \label{eq:casreal}
\eee 
where $\chi_V(c)$ denotes the trace of $c$ acting in $V$, which does not depend on the choice of $c \in [c],$ and $|[c]|$ is the number of elements in the conjugacy class $[c]$.

We say that a Casimir element is \textit{real} if it acts as multiplication by a real number on each irreducible representation of $G$.  
For the Casimir elements \eqref{eq:cassum} this is equivalent to the condition $\chi_V(c) \in \mathbb{R}$ for each  irreducible representation $V$ of $G$.

\begin{Lemma} \label{Lemma:realcas}
Let $C$ be a real Casimir element of $G$ of the form \eqref{eq:cassum} for a conjugacy class $[c]$.  Then for any $c \in [c]$ the inverse $c^{-1}$ belongs to $[c]$ as well.  
\end{Lemma}
\begin{Proof}
For any representation of a finite group $G$ and any $g \in G$ we have 
\be
\chi(g^{-1}) = \overline{\chi(g)}. 
\ee
Therefore, if the character is real then $\chi(c^{-1}) = \chi(c)$.  Since the conjugacy classes are distinguished by their characters this means that $c$ and $c^{-1}$ belong to the same conjugacy class.     \qed
\end{Proof}

Given a finite group $G$ and a subgroup $H \subset G$, we can use a real Casimir element $C$ of the form \eqref{eq:cassum} 
to construct a graph $\Gamma_C=\Gamma(G,H,C)$ as follows.  The set of vertices $\mathcal V$ of $\Gamma_C$ are defined to be the left cosets $G/H$. If $x,y$ are two distinct cosets of $G/H$, we draw an edge  from $x$ to $y$ if there exists a summand $c$ of $C$ such that $c \cdot x = y$.  We do not draw  edges starting and ending at the same point and we do not draw multiple edges between different points.  
Note that from Lemma \ref{Lemma:realcas} it follows that that for any edge from $x$ to $y$ there exists an edge from $y$ to $x$, so as a result we have the undirected graph with simple edges.

\begin{Lemma} \label{Lemma:dreg}
The graph $\Gamma_C$ is $d$-regular for some non-negative integer $d$.  
\end{Lemma}
\begin{Proof}
We have to show that if there are $d$ edges connected to $x\in G/H$ then there are $d$ edges connected to any other $y \in G/H$.  From the transitivity of $G$-action on $G/H$ there exists $g \in G$ such that $g \cdot x = y$. Let $[x,x_1], \ldots, [x,x_d]$ be the edges connected to vertex $x$ with $c_1, \ldots, c_d \in [c]$ such that $x_i=c_i \cdot x.$ Then there are $d$ edges $[y, y_i], \, y_i=g \cdot x_i$ since $y_i= g c_i \cdot x= g c_i g^{-1} \cdot y$  and $g c_i g^{-1} \in [c]$.  \qed
\end{Proof}

We say that $(G,H,C)$ is a \textit{good triple} if the group $G$ acts transitively also on the set of edges of $\Gamma_C$.   
We will show now that under this assumption the adjacency matrix and the Laplacian for $\Gamma_C$ can be described in terms of {\it induced representations}, already mentioned in Section 2 above. In fact, the example of Dirac magnetic monopole from Section 2 was the main motivation for our construction.

Induced representations were first defined by Frobenius in the context of finite groups. We give a brief account of them here, referring for details to \cite{FH}.   

Given any group $G$ and its subgroup $H$, one can readily define a representation $\res_G^H(V)$ of $H$ from any representation $V$ of $G$ by restriction.  Induction is a way of going in the other direction, namely creating from a representation $W$ of $H$ a new representation $\ind_H^G(W)$ of $G$.  
For Lie groups, the representation space of the induced representation has a direct geometric interpretation as the space of sections of a vector bundle over the homogeneous space $G/H$ with fibre $W$, with the group $G$ acting by translations on the base (see \cite{Bott} and the example in Section 2).      

For finite groups the situation is consistent, but necessarily more formulaic.  For each coset $x \in G/H$, choose a representative $g_x$ (the choice turns out not to matter) and take a copy $W_x$ of $W$.  For $w \in W$, denote by $g_x w$ the corresponding element in $W_x$.  
The induced representation $\ind_H^G(W)$ is then formed by taking the direct sum of all these copies of $W$  
\be
\ind_H^G(W) := \bigoplus_{x \in G/H} W_x.
\ee
Any element $v$ of $\ind_H^G(W)$ may be written as $v = \sum{g_x w_x}, \, w_x \in W$.  The action of the group $G$ on this space is defined by the formula
\bee
g\cdot(g_x w_x) = g_y (h \cdot w_x) \qquad \text{if} \ \ g g_x = g_y h. \label{eq:gaction}
\eee
One can check that this indeed gives a representation of $G$.

The induced representation of $G$ is not, in general, irreducible.  Indeed, it may be decomposed into irreducible representations of $G$ according to the Frobenius Reciprocity Theorem \eqref{eq:frobG}, which shows that restriction and induction are adjoint functors.  For finite groups the decomposition of the induced representation may be computed easily using the character tables of $G$ and $H$.  

Now we explain how to construct the magnetic Laplacian from a good triple $(G,H,C)$ and a character $\rho: H \rightarrow U(1).$ 

Suppose that $[x,y]$ is an edge of $\Gamma_C$, then there exist  $c_1, \ldots, c_l \in [c]$, such that $c_i \cdot x = y$.  In terms of coset representatives $g_x$ of $x$ and $g_y$ of $y$, this means that
\bee
c_i g_x = g_y h_i \label{eq:cosrep}
\eee
for some $h_i \in H$.  We associate to the edge $[x,y]$ the element of $U(1)$ given by
\bee
\exp[i \theta_{xy}] = \frac{\rho(h_1) + \ldots + \rho(h_l)}{|\rho(h_1) + \ldots + \rho(h_l)|} \label{eq:magf}
\eee
assuming that the condition 
\bee
\rho(h_1) + \ldots + \rho(h_l) \neq 0 \label{eq:degenmag}
\eee
is satisfied. For a good triple $(G,H,C)$ the validity of \eqref{eq:degenmag} for one edge implies its validity for all edges by the transitivity of $G$-action on the edges of $\Gamma_C.$

\begin{Remark}  It may be the case that the condition \eqref{eq:degenmag} holds only for certain characters of $H$ whilst failing for others.  In such a case sometimes one can still define a magnetic field for the failing case using a non-failing one.  In particular, for the icosahedral group considered in the next section, this condition fails for the case with Chern number $\pm 5$, however, by raising to the power 5 each matrix element of the magnetic adjacency matrix with Chern number $\pm1$  we get the matrix $A_I^5$.  This corresponds in the continuous case to computing the connection form on a principal bundle and then using this to compute the connection form on an associated bundle using the matrix representation. \end{Remark}

\begin{Lemma}
This  construction does indeed define a magnetic potential $\mc{A}_\rho$ on $\Gamma_C$, so we have $\exp[i \theta_{xy}] = \exp[-i \theta_{yx}]$. 
\end{Lemma}
\begin{Proof}
This follows from the reality of $C$.  Indeed, from Lemma \ref{Lemma:realcas} in that case if $c$ is a summand of $C$, then so is $c^{-1}$.  Rewriting \eqref{eq:cosrep}, we see that
\be
c_i^{-1}  g_y = g_x  h_i^{-1}
\ee
and thus 
\bee
\exp[i \theta_{yx}] = \frac{\rho(h^{-1}_1) + \ldots + \rho(h^{-1}_l)}{|\rho(h^{-1}_1) + \ldots + \rho(h^{-1}_l)|} = \frac{\overline{\rho(h_1)} + \ldots + \overline{\rho(h_l)}}{|\rho(h_1) + \ldots + \rho(h_l)|} = \exp[-i \theta_{xy}]. \label{eq:magf2}
\eee
\qed
\end{Proof}

\begin{Proposition} \label{proposition:reptheorymagadj}
If $(G,H,C)$ is a good triple then the magnetic adjacency matrix $A_\rho$ for the magnetic field $\mc{A}_\rho$ on $\Gamma_C$ is given by
\bee
A_\rho = \frac{1}{p}\left(C_\rho - q I\right), \label{eq:magadjgammac}
\eee
where $C_\rho$ denotes the matrix of $C$ in  $\ind_H^G(\rho)$ and $p$ and $q$ are some real constants. The corresponding magnetic Laplacian is given by
\bee
\mathcal L_\rho = d I - A_\rho \label{eq:maglapgammac}
\eee 
since $\Gamma_C$ is $d$-regular.  
\end{Proposition}
\begin{Proof}
We want to show that the matrix $C_\rho$ has a constant $q$ along the diagonal and that all  non-zero off-diagonal terms have constant modulus $p$.  

The first part follows from the fact that $G$ acts transitively on $G/H$.  Indeed, suppose that $x,y$ are two cosets such that $y = g \cdot x$ and that there 
are $m$ summands $c_1, \ldots, c_m \in [c]$  such that $c_i \cdot x =x$. Then there are $m$ summands such that $c'_i \cdot y=y$ given by $c'_1 := g c_1 g^{-1}, \ldots, c'_m := g c_m g^{-1} \in [c]$.   On the level of coset representatives, this means that given $g  g_x = g_y  h$ and $c_i  g_x = g_x h_i$  we have
\be
c'_i g_y = g_y h h_i h^{-1}
\ee
and thus 
\be
(C_\rho)_{yy} = \rho(h h_1 h^{-1}) + \ldots + \rho(h h_m h^{-1}) = \rho(h_1) + \ldots + \rho(h_m) = (C_\rho)_{xx}=q \in \mbb{R}.
\ee
The reality condition follows from the fact that if $c \cdot x = x$ then $c^{-1} \cdot x = x$.  

The second part follows from the assumption of the transitivity of $G$-action on the edges of $\Gamma_C$.  Indeed, for any edges $[x,y]$ and $[x',y']$, there exists $g \in G$ such that $g \cdot [x,y] = [x',y']$.  On the group level we have that $g g_x = g_{x'} h_x$ and $g g_y = g_{y'} h_y$.  Given $l$ elements $c_1, \ldots, c_l$ such that $c_i \cdot x = y$ there are $l$ elements $c'_1 = g c_1 g^{-1}, \ldots, c'_l= g c_l g^{-1}$ such that $c'_i \cdot x' = y'$, and thus
\be
c_i g_x = g_y h_i \ \ \text{and} \ \ c'_i g_{x'} = g_{y'} h_y h_i h_x^{-1}.
\ee
Computing $(C_\rho)_{xy}$ and $(C_\rho)_{x'y'}$ we have that
\be
(C_\rho)_{xy} = \rho(h_1) + \ldots \rho(h_l) 
\ee
and 
\be
(C_\rho)_{x'y'} = \rho(h_y h_1 h_x^{-1}) + \ldots \rho(h_y h_l h_x^{-1}) = \rho(h_y) (C_\rho)_{xy} \rho(h_x^{-1}).
\ee  
Therefore we have that $|(C_\rho)_{x'y'}| = |(C_\rho)_{xy}|=p \in \mbb{R}$ and hence \eqref{eq:magadjgammac} and \eqref{eq:maglapgammac}.       \qed
\end{Proof}

\begin{Proposition} \label{Proposition:fluxface}
The magnetic field defined by \eqref{eq:magf} is $G$-invariant in the following sense: if $\gamma$ is a cycle on $\Gamma_C$ and $\gamma' = g \cdot \gamma$ then the fluxes through the cycles $\gamma$ and $\gamma'$ are the same.     
\end{Proposition}
\begin{Proof}
This follows by iteratively applying the argument from the second half of the proof of Proposition \ref{proposition:reptheorymagadj} to each edge in a cycle.  \qed
\end{Proof}

\begin{Theorem} \label{Theorem:spec}
If $(G,H,C)$ is a good triple then the spectrum of the magnetic adjacency matrix and magnetic Laplacian on the graph $\Gamma_C$ defined above can be found explicitly using the character tables of $G$ and $H$ and the Frobenius reciprocity formula.  
\end{Theorem}
\begin{Proof}
Using Frobenius reciprocity theorem and the character tables of the groups $G$ and $H$ we can decompose induced representations into irreducible representations of $G$.  Specifically, for $\rho$ a character of $H$ and $V_i$ distinct irreducible representations of $G$, we have a decomposition of the form 
\be
\ind_H^G(\rho) \cong V_1^{\oplus_{i_1}} \oplus \ldots V_r^{\oplus_{i_r}}
\ee
where $\sum_{k=1}^r{i_k \cdot \dim(V_k)} = |G/H|$.  The spectrum of $C_\rho$ is then given by computing the eigenvalues of $C_\rho$ acting in each $V_k$ using formula \eqref{eq:casreal}, with the multiplicity of each eigenvalue $\lambda_k$ is given by $i_k \cdot \dim(V_k)$: 
\be
Spec(C_\rho) = \left\{[\lambda_{1}]^{i_1 \cdot \dim(V_1)}, \ldots,[\lambda_{r}]^{i_k \cdot \dim(V_k)} \right\}.
\ee  
Consequently, the spectrum of the magnetic adjacency and Laplacian matrices are given by using \eqref{eq:magadjgammac} and \eqref{eq:maglapgammac} to shift and scale the spectrum of $C_\rho$. \qed
\end{Proof}

\section{Binary polyhedral groups and Dirac's monopoles on polyhedral graphs}

Since we are looking for a discrete analogues of Dirac's monopole on $S^2$ we have to add now the condition that the corresponding graph $\Gamma$ is {\it planar.}
For planar graphs we have the notion of faces $F \in \mathcal F(\Gamma)$ and can define an analogue of the {\it Chern number} by summing the magnetic fluxes \eqref{eq:mflux} through each face $F$ taken with orientation:
\bee
c(\mc{A}) = \frac{1}{2 \pi}\sum_{F \in \mathcal F(\Gamma)}{\Phi(F)}. \label{eq:discchern}
\eee
It is clear that this quantity is only defined up to a sign depending on a choice of orientation. Recall that in the continuous case the spectrum \eqref{eq:spec} is symmetric under $q \mapsto -q$.  The same is true here: given a magnetic potential $\mc{A}$, replacing every phase by its complex conjugate just amounts to replacing $A_{\mc{A}}$ and $\mathcal L_{\mc{A}}$ by their transposes. 

The most natural discrete versions of $S^2$ are the polyhedral graphs of Platonic solids. 
It is very natural therefore to apply our construction to the class of {\it binary polyhedral groups} $2T, 2O, 2I \subset SU(2)$ going back to Klein \cite{Klein}.

We will give now the examples of analogues of Dirac magnetic monopoles for all Platonic solids with the exception of dodecahedron, which are coming from our construction.  In each case, we will explain what the choices of $G,H$ and $C$ are that generate the particular graph and give explicitly the corresponding magnetic spectra. We stress that for each example the magnetic potential is described for a particular choice of coset representatives, which does not matter, but may hide the full symmetry of the system (cf. usual Dirac magnetic monopole on the sphere). The choice of $C$ is the result of experimenting, full details of the calculations may be found in \cite{K}.

\subsection{Dirac magnetic monopoles on the tetrahedron}

The group of orientation-preserving symmetries  of the tetrahedron $T\subset SO(3)$ can be naturally identified with the alternating group $A_4$, which has 12 elements; whilst the stabilizer of a vertex is a cyclic group $\mbb{Z}_3$ of order $3$.   

Under the double-covering $SU(2) \rightarrow SO(3)$ these lift to the {\it binary tetrahedral group} $G=2T \subset SU(2)$, which has 24 elements and its subgroup $H \cong \mbb{Z}_6$.  There are thus six distinct characters of $H$, given by sending the generator to the different sixth-roots of unity.  

\begin{figure}[ht]
	\centering
		\includegraphics{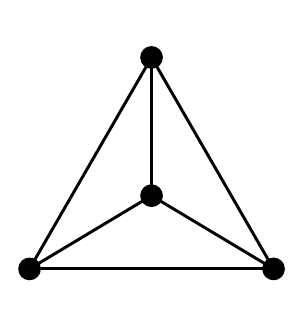}
		\caption{The graph of the tetrahedron. }
		\label{Figure:tet}
\end{figure}

In order to get the tetrahedral graph we take as the Casimir element $C$ the sum of eight elements coming from the conjugacy classes of $(123)$ and $(132)$ in $A_4$.  Acting on the coset space, the summands map each coset to each other coset twice and to itself twice.  According to the definition of the graph $\Gamma_C=\Gamma_C(G,H)$ given in the previous Section we obtain the tetrahedral graph, shown in Figure \ref{Figure:tet}.   

For some choice of coset representatives and ordering of the vertices, we arrive at the following magnetic adjacency matrices for the tetrahedral graph 
\be A_k^T = 
\left( \begin{array}{cccc} 
0	& i^k & i^k & i^k  \\
(-i)^k	& 0 & i^k & (-i)^k \\
(-i)^k & (-i)^k & 0 & i^k  \\
(-i)^k & i^k & (-i)^k & 0 \\
\end{array} \right),
\ee
where $k=-1,0,1,2$. This describes four distinct invariant magnetic fields on the tetrahedral graph.  Notice that $k=0$ corresponds to the adjacency matrix for the tetrahedral graph and $k=2$ corresponds to $-1$ times the  adjacency matrix for the tetrahedral graph.  The matrices $A_k^T$ for $k=1$ and $k=-1$ are complex conjugate.  

The Chern number for this magnetic field described by $A_k^T$ may be computed using \eqref{eq:discchern} and we find that $c(A_k)=k$, for $k=0, \pm 1, 2$, since the flux through each triangular face is $\pm \pi k/2$.  


Since the tetrahedral graph is $3$-regular, the corresponding magnetic Laplacian is given by 
\be
\mathcal L^T_k=3I_4-A^T_k,
\ee 
where $I_4$ is the $4 \times 4$ identity matrix.  

On computing the decomposition of the induced representations into irreducible representations of $2T$ and computing the eigenvalues of $C$ acting in each, we can immediately compute the eigenvalues of the magnetic adjacency matrix and of the magnetic Laplacian (which, of course, in this case can be easily computed directly), see Table \ref{Table:spectetstar}.  

\begin{table}[ht]
	\centering
		\begin{tabular}{|c|c|c|}
		\hline
		  
Chern number& Magnetic adjacency spectrum &  Magnetic Laplacian spectrum  \\
			\hline 
$0$		&	$ [3]^1,[-1]^3,$	& $[0]^1, [4]^3$\\
$\pm1$		&	$[\sqrt{3}]^2$,$[ -\sqrt{3}]^2$	& $[3-\sqrt{3}]^2$,$[ 3+\sqrt{3}]^2$	 \\
$\pm2$		&	$[1]^3,[-3]^1$ & $[2]^3,[6]^1$	\\

			\hline
		\end{tabular}
		\bigskip
				\caption{Magnetic spectra for the tetrahedral graph.}
				\label{Table:spectetstar}
\end{table}

We note that the degeneracy of the ground state of the magnetic Laplacian increases with the strength of the magnetic field, as does the energy level of the ground state --- just as in the continuous case.

\subsection{Dirac magnetic monopoles on the octahedron}

The rotational symmetry group of the octahedron $O\subset SO(3)$ has 24 elements and can be identified with the symmetric group $S_4$.  The stabilizer of a vertex is a cyclic group $\mbb{Z}_4$ of order $4$.  

The group $G$ in this case is the {\it binary octahedral group} $2O \subset SU(2)$ has 48 elements and contains $2T$ as a subgroup (reflecting the fact that one may inscribe two tetrahedra in an octahedron).  The stabilizer of a vertex lifts to $H \cong \mbb{Z}_8$, which means that there are 8 distinct characters corresponding to the different eighth-roots of unity. 

\begin{figure}[ht]
	\centering
		\includegraphics{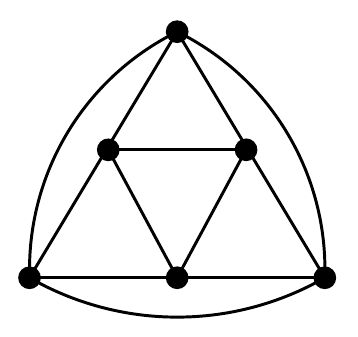}
		\caption{The graph of the octahedron. }
		\label{Figure:octahedron}
\end{figure}

As the Casimir element $C$ we choose the sum of the 6 elements in the conjugacy class of $(1234)\in S_4$.  Acting on $G/H$, they map each coset to itself twice and to four other cosets once.   Thus we arrive at the octahedral graph shown in Figure \ref{Figure:octahedron}.

Each different character of $H \cong \mbb{Z}_8$ gives one of the following magnetic adjacency matrices for the octahedral graph
\be A^O_k = 
\left( \begin{array}{cccccc} 
0	& \zeta^k & \zeta^{3k} & 0  &\zeta^{5k} &\zeta^{3k}   \\
\zeta^{7k}  & 0 & \zeta^k & \zeta^{7k}& 0 & \zeta^{3k} \\
\zeta^{5k}	 &  \zeta^{7k}& 0 &\zeta^{7k}& \zeta^k& 0 \\
0	 & \zeta^k & \zeta^{k} & 0 & \zeta^k & \zeta^{5k}  \\
\zeta^{3k}  & 0 & \zeta^{7k} & \zeta^{7k}& 0 & \zeta^{5k} \\
\zeta^{5k}	 & \zeta^{5k} &0 & \zeta^{3k} &\zeta^{3k} & 0 \\
\end{array} \right),
\ee
where $\zeta = \exp[2 \pi i/8]$ and $k=-3,-2,\ldots,3,4$ --- with $k=0$ corresponding to the usual adjacency matrix of the octahedral graph.    

For each $k$, we find that the corresponding flux through each triangular face is given by $\pm 2 \pi k /8$ (depending on how the graph is drawn) and since there are 8 faces, by \eqref{eq:discchern} the Chern number is $\pm k$.  

Since the octahedral graph is $4$-regular, the corresponding magnetic Laplacian is given by 
\be
\mathcal L_k^O = 4I_6-A^O_k,
\ee
where $I_6$ is the $6 \times 6$ identity matrix.  

The eigenvalues for the magnetic adjacency and Laplacian matrices are given in Table \ref{Table:specoctstar}.  We note that again the degeneracy of the ground state of the magnetic Laplacian increases with the Chern number up to $c=4$, as does the energy level of the ground state.  

\begin{table}[ht]
	\centering
		\begin{tabular}{|c|c|c|}
		\hline
		  
			 	Chern number& Magnetic adjacency spectrum & Magnetic Laplacian spectrum \\
			\hline 
		0		&	$[4]^1, [0]^3, [-2]^2$ & $[0]^1,[4]^3, [6]^2$	\\
	$\pm 1$		&	$[2\sqrt{2}]^2$,$[ -\sqrt{2}]^4$ & $[4-2\sqrt{2}]^2$,$[ 4+\sqrt{2}]^4$ 	\\
	$\pm 2$		&	$[2]^3,[-2]^3$	& $[2]^3,[6]^3$  \\
	$	\pm 3$		&	$[\sqrt{2}]^4,[-2 \sqrt{2}]^2$ & $[4- \sqrt{2}]^4,[4+2\sqrt{2}]^2$	\\
	$\pm 4$		&	$[2]^2$,$[0]^3$,$[-4]^1$ & $[2]^2,[4]^3,[8]^1$	\\
			\hline
		\end{tabular}
		\bigskip
				\caption{Magnetic spectra for the octahedral graph.}
				\label{Table:specoctstar}
\end{table}

\subsection{Dirac magnetic monopoles on the cube}

The rotational symmetry group of the cube is the same as that of the octahedron, but the stabilizer of a vertex is a cyclic group $\mbb{Z}_3$ of order $3$.  Thus, in this example $G=2O$ is the binary octahedral group and $H=\mbb{Z}_6$ is a cyclic subgroup of order 6 with 6 distinct characters corresponding to different sixth-roots of unity.    

\begin{figure}[h]
\centerline{ \includegraphics[width=4cm]{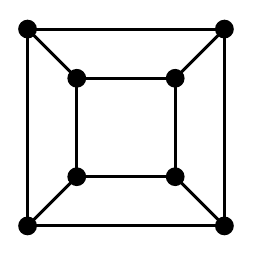} }
		\caption{The graph of the cube. }
		\label{Figure:cube}
\end{figure}

As $C$ we choose the same Casimir element as in the octahedral case. This will lead to the cubic graph, shown in Figure \ref{Figure:cube}.

For some ordering of the vertices we have the following magnetic adjacency matrices
\begin{equation}
 A^C_k = 
\left( \begin{array}{cccccccc} 
0 & 0 & 0 & 0 & \theta^{11k} & \theta^{7k} & 0 & \theta^{9k}\\
0 & 0 & 0 & 0 & \theta^k & \theta^{11k} &  \theta^{9k}  & 0\\
0 & 0 & 0 & 0 & \theta^{3k} & 0 & \theta^k & \theta^{11k}\\
0 & 0 & 0 & 0 & 0 & \theta^{9k} & \theta^{5k} & \theta^k\\
\theta^k & \theta^{11k} & \theta^{9k} & 0 & 0 & 0 & 0 & 0\\
\theta^{5k} & \theta^k & 0 & \theta^{3k} & 0 & 0 & 0 & 0\\
0 & \theta^{3k} & \theta^{11k} & \theta^{7k} & 0 & 0 & 0 & 0\\
\theta^{3k}& 0 & \theta^k & \theta^{11k} & 0 & 0 & 0 & 0
\end{array} \right),
\end{equation}
where $\theta = \exp[2 \pi i /12]$ and $k=-2,-1,0,1,2,3$.  

For each $k$, we find that the corresponding flux through each triangular face is given by $\pm k 2 \pi/8$ (depending on the orientation) and since there are 8 faces, by \eqref{eq:discchern} the Chern number is $\pm k$.  

Since the cubic graph is $3$-regular, the corresponding magnetic Laplacian is given by 
\be
\mathcal  L_k^C = 3I_8-A^C_k,
\ee
where $I_8$ is the $8 \times 8$ identity matrix.  The eigenvalues for the magnetic adjacency and Laplacian matrices are given in Table \ref{Table:speccubestar}.  Note that the degeneracy of the ground state of the magnetic Laplacian again increases with the Chern number.  

\begin{table}[ht]
	\centering
		\begin{tabular}{|c|c|c|}
		\hline
		  
			 	Chern number& Magnetic adjacency spectrum & Magnetic Laplacian spectrum \\
			\hline 
		0		&	$[3]^1, [1]^3, [-1]^3,[-3]^1$ & $[0]^1,[2]^3,[4]^3, [6]^1$	\\
	$\pm 1$		&	$[\sqrt{6}]^2$,$[0]^4$,$[ -\sqrt{6}]^2$ & $[3-\sqrt{6}]^2$,$3^4$,$[ 3+\sqrt{6}]^2$ 	\\
	$\pm 2$		&	$[2]^3,[0]^2,[-2]^3$	& $[1]^3,[3]^2,[4]^3$  \\
	$	\pm 3$		&	$[\sqrt{3}]^4,[- \sqrt{3}]^4$ & $[3- \sqrt{3}]^4,[3+\sqrt{3}]^4$	\\
			\hline
		\end{tabular}
		\bigskip
				\caption{Magnetic spectra for the cubic graph.}
				\label{Table:speccubestar}
\end{table}

\subsection{Dirac magnetic monopoles on icosahedron}

The rotational symmetry group of the icosahedron $I \subset SO(3)$ has 60 elements and can be identified with the alternating group $A_5 \subset S_5$.  This can be seen by noting that there are five inscribed tetrahedra with vertices at the midpoint of each face of the icosahedron and rotations give any even permutations of these.  
The stabilizer of a vertex is a cyclic group $\mbb{Z}_5$ of order $5$.  

The group $G$ in this case is the {\it binary icosahedral group} $2I \subset SU(2)$ has 120 elements and the subgroup $H \cong \mbb{Z}_{10}$ is a cyclic subgroup of order 10.  There are 10 different characters of $H$, corresponding to the different tenth roots of unity. 

\begin{figure}[ht]
	\centering
		\includegraphics{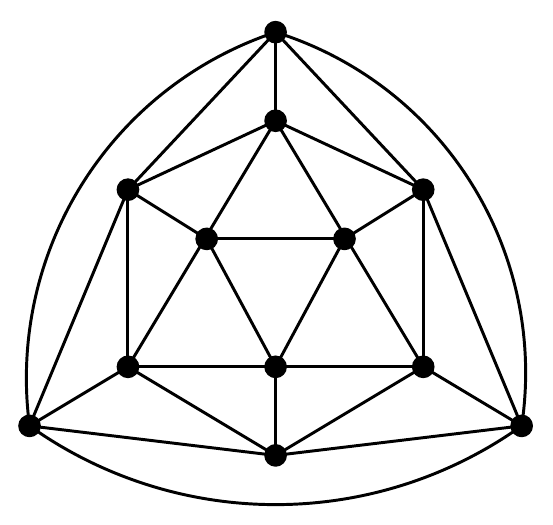}
		\caption{The graph of the icosahedron. }
		\label{Figure:icos}
\end{figure}

To arrive at the right graph we choose as $C$ the sum of the 12 elements corresponding to the conjugacy class of $(12345)$.  (Recall that the elements $(12345)$ and $(21345)$ are not conjugate in $A_5$, only in $S_5$.)  Acting on the coset space $G/H$ , we find  that two elements map each coset to itself, whilst the remaining ten elements are distributed evenly between five cosets.  This generates the icosahedral graph, shown in Figure \ref{Figure:icos}.    

Let $\eta := \exp[2 \pi i /20]$, then for $k=-9,-8,\ldots, 9, 10$ we have the following set of magnetic adjacency matrices for the icosahedral graph (for a suitable labelling of the vertices) 
\be  
A^k_I=\left( \begin{array}{cccccccccccc} 
0& 0 & \eta^{5k} &0 & \eta^{9k}& \eta^k & 0 & 0  & \eta^{17k} & \eta^{13k} & 0 & 0 \\
0& 0 & 0 &\eta^{15k} & 0& 0 & \eta^{19k} & \eta^{11k}  & 0 & 0 & \eta^{7k} & \eta^{3k} \\
\eta^{15k}& 0 & 0 &0 & 0& 0 & \eta^{15k} & \eta^{5k}  & \eta^{11k} & \eta^{9k} & 0 & 0 \\
0& \eta^{5k} & 0 &0 & \eta^{5k}& \eta^{15k} & 0 & 0  & 0 & 0 & \eta^{11k} & \eta^{9k} \\
\eta^{11k}& 0 & 0 &\eta^{15k} & 0& \eta^{11k} & 0 & 0  & \eta^{9k} & 0 & \eta^{5k} & 0 \\
\eta^{19k}& 0 & 0 &\eta^{5k} & \eta^{9k}& 0 & 0 & 0  & 0 & \eta^{11k} & 0 & \eta^{15k} \\
0& \eta^k & \eta^{5k} &0 & 0& 0 & 0 & \eta^{11k}  & \eta^{15k} & 0 & \eta^{9k} & 0 \\
0& \eta^{9k} & \eta^{15k} &0 & 0& 0 & \eta^{9k} & 0  & 0 & \eta^{5k} & 0 & \eta^{11k} \\
\eta^{3k}& 0 & \eta^{9k} &0 & \eta^{11k}& 0 & \eta^{5k} & 0  & 0 & 0 & \eta^{15k} & 0 \\
\eta^{7k}& 0 & \eta^{11k} &0 & 0& \eta^{9k} & 0 & \eta^{15k}  & 0 & 0 & 0 & \eta^{5k} \\
0& \eta^{13k} & 0 &\eta^{9k} & \eta^{15k}& 0 & \eta^{11k} & 0  & \eta^{5k} & 0 & 0 & 0 \\
0& \eta^{17k} & 0 &\eta^{11k} & 0& \eta^{5k} & 0 & \eta^{9k}  & 0 & \eta^{15k} & 0 & 0 \\
 \end{array} \right) 
\ee

We find that, depending on the orientation, the flux through each triangular face is given by $\pm 2\pi k/20$.  Since there are 20 faces of the icosahedron, by \eqref{eq:discchern} we see that the Chern number is $\pm k$.  


Since the icosahedral graph is $5$-regular, the corresponding magnetic Laplacian is given by $
\mathcal L_I^k = 5I_{12} - A_I^k.$
The eigenvalues of the magnetic adjacency matrices and magnetic Laplacians are easily computed using the character tables of the group and are shown in Table \ref{Table:specicosstar}.  

We see again that the ground state of the magnetic Laplacian increases in degeneracy with the modulus of the Chern number, as does the energy level of the ground state.  In this example and the previous one, the pattern stopped at some point as the Chern number increased. It is interesting to note that one can formally continue this pattern by grouping all the eigenvalues except the last one --- in which case the ``ground state" increases in degeneracy and the average ``ground state" also increases.  

\begin{table}[ht]
	\centering
		\begin{tabular}{|c|c|}
		\hline
		  
		Chern number& Magnetic adjacency spectrum   \\
		\hline
		$0$ 			& $[-\sqrt{5}]^3, [-1]^5,  [\sqrt{5}]^3,[5]^1  $	 		\\

		$\pm 1$ 	& $\left[-\sqrt{\frac{1}{2}(5+ \sqrt{5})}\right]^6 , \left[\sqrt{5-2\sqrt{5}}\right]^4, \left[\sqrt{\frac{5}{2}(5+ \sqrt{5})}\right]^2$ 	 \\

		$\pm 2$ 	&$ \left[-\sqrt{5}\right]^4,\left[\frac{1}{2}\left(-3+\sqrt{5}\right)\right]^5,\left[\frac{1}{2}\left(5+ \sqrt{5} \right)\right]^3 $ 	\\

		$\pm 3$ 	& $\left[-\sqrt{\frac{5}{2}(5 - \sqrt{5})}\right]^2 , \left[-\sqrt{\frac{1}{2}(5 - \sqrt{5})}\right]^6, \left[\sqrt{5+2\sqrt{5}}\right]^4 $	\\

		$\pm 4$ 	&$\left[-\sqrt{5}\right]^4 ,\left[\frac{1}{2}\left(-5+\sqrt{5}\right)\right]^3,\left[\frac{1}{2}\left(3+ \sqrt{5} \right)\right]^5  $ 	\\

		$\pm 5$ 	& $\left[-\sqrt{5}\right]^6, \left[\sqrt{5}\right]^6$	\\

		$\pm 6$ 	&$\left[\frac{-1}{2}\left(3+ \sqrt{5} \right)\right]^5, \left[\frac{1}{2}\left(5-\sqrt{5}\right)\right]^3,\left[\sqrt{5}\right]^4 $	\\

		$\pm 7$ 	& $\left[-\sqrt{5+2\sqrt{5}}\right]^4, \left[\sqrt{\frac{1}{2}(5 - \sqrt{5})}\right]^6,\left[\sqrt{\frac{5}{2}(5 - \sqrt{5})}\right]^2$ \\	

		$\pm 8$ 	& $\left[\frac{-1}{2}\left(5+ \sqrt{5} \right)\right]^3,\left[\frac{1}{2}\left(3-\sqrt{5}\right)\right]^5,\left[\sqrt{5}\right]^4  $	\\

		$\pm 9$ 	& $\left[-\sqrt{\frac{5}{2}(5+ \sqrt{5})}\right]^2,\left[-\sqrt{5-2\sqrt{5}}\right]^4,\left[\sqrt{\frac{1}{2}(5+ \sqrt{5})}\right]^6$  	\\

		$\pm 10$	& $[-5]^1,[-\sqrt{5}]^3,[1]^5,[\sqrt{5}]^3 $ \\

			\hline
		\end{tabular}
		\bigskip
						\caption{Eigenvalues for the magnetic adjacency matrix for the icosahedral graph.  The spectrum of the magnetic Laplacian can be obtained from the relation $\lambda_{\mathcal L} = 5-\lambda_A$.  }
				\label{Table:specicosstar}
\end{table}

\begin{figure}[h]
\centerline{ \includegraphics[width=10cm]{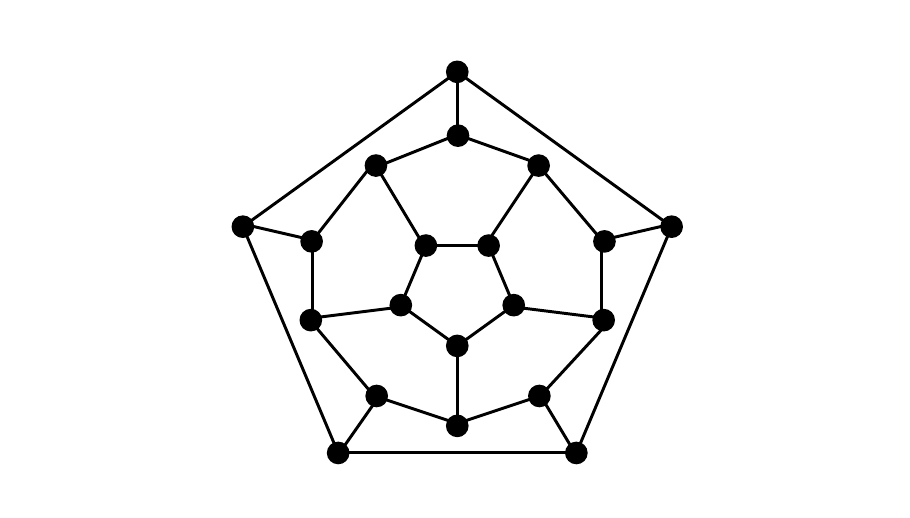} }
		\caption{The graph of the dodecahedron. }
		\label{Figure:dodec}
\end{figure}

In the remaining dodecahedral case (see Figure \ref{Figure:dodec}) $G$ is again the binary icosahedral group $2I$ with the subgroup $H=\mbb{Z}_6$.
However, a suitable Casimir element $C$ does not exist.

\begin{Proposition} \label{Proposition:dodec}
For $G=2I$ and $H=\mbb{Z}_6$ there are no Casimir elements $C$, which produce the dodecahedral graph by the construction described above.    
\end{Proposition}
\begin{Proof}
It is known (see e.g. \cite{ST}) that the eigenvalues of the adjacency matrix of the dodecahedral graph (with multiplicities) are
$$[-\sqrt{5}]^3, \, [-2]^4, \, [0]^4, \, [1]^5, \, [\sqrt{5}]^3,\, [3]^1.$$
On the other hand, one can check that the space $C(\mathcal V)$ of functions on the vertices of dodecahedron as a $G$-module 
has the decomposition into irreducible modules with two copies of the same irreducible representation of dimension 4 appearing twice.
This means that if the required Casimir element $C$ would exist, then it would have on $C(\mathcal V)$ an eigenvalue of multiplicity at least 8.
By \eqref{eq:magadjgammac} the same must be true for the adjacency matrix as well. This contradiction means that no such $C$ can exist. \qed
\end{Proof}

This means that probably we should extend our construction, e.g. by relaxing the transitivity on edges condition and allowing multiple edges.  This will allow to use any real Casimir element $C$ by drawing an edge for the action of every summand of $C$. 

\begin{figure}[h]
\centerline{ \includegraphics[width=6cm]{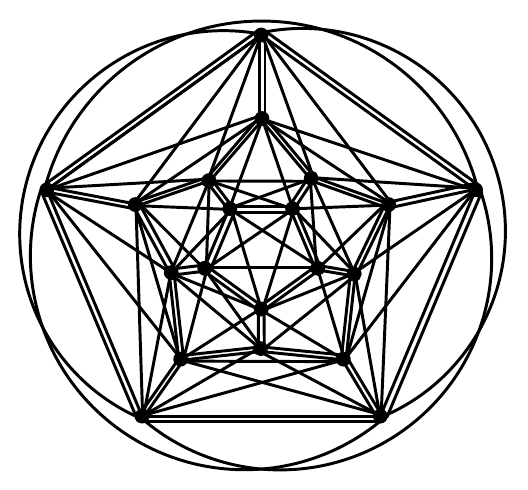} }
		\caption{"Pentadodecahedral" graph.}
		\label{Figure:pentadod}
\end{figure}

For instance, if we do this for $G=2I, \, H=\mbb{Z}_6$ and the Casimir $C$ that was used in the construction of the icosahedral graph, then we arrive at a non-planar cousin $\Gamma$ of the dodecahedral graph, which contains both single and double edges. It is formed by glueing 12 copies of pentagrams $K_5$ (in contrast to pentagons in the dodecahedral case) along the double edges, see Fig. \ref{Figure:pentadod}.

The spectrum of the corresponding Laplacian is 
\be 
[0]^1, [10-2\sqrt{5}]^3, [12]^5,[10+2\sqrt{5}]^3,[15]^8.
\ee
Note that the two copies of the irreducible modules of dimension 4 
in the space $C(\mathcal V)$ now have the same eigenvalue 15 with total multiplicity 8.
Since the graph $G$ could be naturally embedded to the surface of genus 12, 
which is a result of gluing a handle to each face of dodecahedron, 
the corresponding magnetic Laplacian can be considered 
as a discrete version of the Landau problem on that surface.

\section{Conclusions}

The method presented here for constructing invariant magnetic fields on the graphs of regular polyhedra is algebraic in nature and is by no means the only possible approach. The failure in the dodecahedral case shows that there are discrete analogues of the Dirac magnetic monopole outside of our scheme.

The main advantage of our method is that we can compute the spectrum of the corresponding magnetic adjacency and Laplacian matrices very easily.  In addition,  our method is natural ---  up to a choice of coset representatives (which amounts to a choice of gauge) everything is canonical.  Furthermore, as much as is possible we have kept the structure of the original Lie group case \cite{KV}. 

From this perspective one can view our construction as an attempt to define a {\it finite group analogue of the coadjoint orbits} of compact Lie groups.
The Casimir element $C$ corresponds to the choice of the normal metric on such an orbit. 

One can extend our approach to the case of infinite discrete groups $G,$ but if the subgroup $H$ is not cofinite we need to use the analytic tools to investigate the spectra. A natural example would be a discrete version of the Landau problem on the regular hexagonal and triangular planar graphs, playing a crucial role, in particular, in S.P. Novikov's approach to the discrete complex analysis. 

It is natural also to look for generalisations of our construction to the non-abelian gauge groups. In this relation we should mention a very interesting work by Manton \cite{Manton}, who discussed the notion of a connection on a discrete fibre bundle, and by Morrison \cite{Morrison}, who described a discrete analogue of the Yang-Mills action for such connections and showed that the examples considered by Manton are minimal in that sense.

\subsection*{Acknowledgements}
We are grateful to Alexey Bolsinov and Derek Harland for very useful discussions.

The work was partially supported by the EPSRC (grant EP/J00488X/1).

\end{document}